# General Theory of Statistical Fluctuations with Applications to Metastable states, Nernst Points, and Compressible Multi-component Mixtures

P. D. Gujrati

Department of Physics, Department of Polymer Science
The University of Akron, Akron, OH 44325

**Abstract**

The general fluctuation theory is reviewed with special attention to the role played by different ensembles, and is extended to incorporate stationary metastable states obtained in the long time limit. The fluctuation in a quantity depends on the nature of the ensemble and contains at most $n$ different fluctuation contributions, where $n \geq 0$ is the number of fluctuating extensive quantities in the ensemble. We prove four general theorems and a corollary for statistical fluctuations valid for any thermodynamic system. We also demonstrate by two examples that the results of the theory remain valid regardless of the magnitude of the fluctuations. To avoid certain physical paradoxes, it is postulated that stationary metastable states like the ideal glass cannot exist in Nature. We also prove a generalized Nernst theorem valid at Nernst points at which certain susceptibility like the heat capacity vanishes. The theorem is no longer restricted to absolute temperature. We calculate statistical fluctuations in the number of monomers and other physical quantities of interest in a compressible mixture. We demonstrate that the density and composition fluctuations are in general not statistically independent, which is contrary to some recent claims. The standard isothermal compressibility at constant monomer numbers does not represent the density fluctuation in all ensembles. We show that the density fluctuation at constant composition is a meaningless concept, except at absolute zero. We prove a relation between the weighted monomer number fluctuation and the volume fluctuation in a multi-component system, which is an extension of a well-known similar relation for a single component system.

## I. Introduction

The present review deals with the general theory of statistical fluctuations[1-7] in statistical mechanics and its applications to pure components and multi-component mixtures. We will also extend this theory to *stationary metastable states* obtained in the long time limit provided the stable phase is forbidden.[7] The *statistical* fluctuation $\Delta Q \equiv Q - \overline{Q}$ in some quantity $Q$ is its deviation from its *statistical* average $\overline{Q} \equiv <Q>$. However, if there is no possibility of confusion, we will omit the bar or the angular bracket and use $Q$ to also denote the average. The theory of fluctuations is a very rich branch of statistical physics and relates fluctuations in $Q$ with an appropriate "susceptibility." These generalized susceptibilities like the specific heat, compressibility, magnetic susceptibility, etc. measure the response of an extensive quantity $X = S, V, M$, etc. to applied intensive field $Y = T, P, H$, etc., given by one of the second derivatives of some thermodynamic potential. Here $S, V, M$, etc. represent the total entropy, the volume, the magnetization, etc., respectively, and $Y = T, P, H$, etc represent the temperature, the pressure, the magnetic field, etc. Provided these susceptibilities are *finite*, usual arguments suggests that any statistical ensemble can be used to describe physical phenomena, as they are all equivalent in thermodynamic limit. Near phase transitions, where some of these susceptibilities become anomalously large, various ensembles may not be equivalent.[1-5] However, the theory of phase transition is well understood,[8] and will not be visited in any detail here. For this, we refer the reader to various volumes in the series in Ref. 8.

There are instances where some susceptibilities *vanish*. For example, the isothermal magnetic susceptibility $\chi_T$ for some superconducting materials can vanish at some intermediate temperatures.[9] Similarly, the specific heats of astrophysical objects can also vanish,[1] especially under gravitational collapse.[10,11] Furthermore, various physical models can be identified as a "non-physical" limit of other physical models,[12] in which susceptibilities can vanish at some finite $Y$.[13,14] Ref. 5 clearly shows that thermodynamics is valid for such unphysical limits. There are also examples in which a susceptibility can vanish in stationary metastable states,[7] like the ideal glass which has zero specific heat below the ideal glass transition[15,16] temperature. Thus, it is also important to study the consequences of vanishing susceptibilities. Some of these consequences turn out to be, indeed, surprising.[5,7] In the following, we will call a point of zero susceptibility a *Nernst point* to distinguish it from a critical point.[5] The most famous and well-known consequence of vanishing specific heat at $T = 0$ is known as the Nernst's theorem.[17]

The issue of statistical fluctuations in the number[1,2] $N_j$ of particles of the *j*-th species in a mixture[18,19] has received considerable interest in recent years,[20-28] chiefly because of the current interest in neutron scattering from blends of hydrogenated and their deuterated counterpart polymers. However, earlier attempts[29-33] should also be mentioned for completeness. The main interest has been focused on a compressible binary mixture because of the contrast produced by deuteration. Thus, we will mostly focus on two species *j*=1 and 2, with $N_{m1}$ and $N_{m2}$ denoting the number of monomers in them, even though the formal analysis will be carried out for multi-component mixtures. We introduce $\mu_{mj}$, *j*=1,2, to denote the chemical potential per monomer. Our interest is in the scattering from the monomers. Therefore, we are interested in the following thermodynamic correlations $<(\Delta N_{m1})^2>$, $<\Delta N_{m1} \Delta N_{m2}>$ and $<(\Delta N_{m2})^2>$ in the *statistical* fluctuations $\Delta N_{m1}$ and $\Delta N_{m2}$ in $N_{m1}$ and $N_{m2}$, among others. In terms of these fluctuations, the total intensity in the forward direction in a scattering experiment is given by[23]

$$I(0|b_1, b_2) \equiv b_1^2 <(\Delta N_{m1})^2> + 2 b_1 b_2 <\Delta N_{m1} \Delta N_{m2}> + b_2^2 <(\Delta N_{m2})^2>, \quad (1)$$

where $b_1$ and $b_2$ are the scattering lengths from the monomers of the two species 1 and 2, respectively. If we set $b_1=b_2=1$, the above intensity reduces to the fluctuation in the total number $N_m \equiv \Delta N_{m1} + \Delta N_{m2}$ of monomers of both species.

$$< (\Delta N_m)^2 > \equiv I(0|1,1) = <(\Delta N_{m1})^2> + 2<\Delta N_{m1}\Delta N_{m2}> + <(\Delta N_{m2})^2>. \qquad (2)$$

It can be shown,[23] see also Secs. VIII and IX below, that the intensity and the total monomer number fluctuation $<(\Delta N_m)^2>$ in the grand canonical ensemble can each be broken into two parts: the K-part, which contains the compressibility $K_{T,N_{m1},N_{m2}}$, and the µ-part, which contains the inverse of the derivative $(\partial \mu_{m1}/\partial N_{m2})_{T,P,N_{m1}}$ as a factor. Our analysis here is prompted by the following two claims in the literature:[22,23,34]

(i) The K-part describes the density fluctuation, and the µ-part describes the composition fluctuation.

(ii) The density and composition fluctuations are statistically *uncorrelated*.

The claims are quite remarkable, if true, since thermodynamics intertwines various fluctuations in extensive quantities; there are usually cross-correlations among them making them *statistically dependent*. The density fluctuation itself gets modified by additional fluctuations, for example, the composition fluctuation and vice versa. We will pay close attention to this issue in this work. Of course, one can envision an ensemble in which there are only composition fluctuations, but no density fluctuations. However, as we will see, such an ensemble is not realistic and does not disprove our claim that the fluctuations are normally correlated. Our critical examination has shown that neither of the two claims can be justified.[27] In this review, we supply the relevant details. Since the scattering intensity has been investigated by us recently elsewhere,[26,28] we will mostly be investigating individual monomer number and the total monomer number fluctuations in this work.

**Scope of the Review**

We are interested in this review in a system that is finite though very large in extent. Thus, we are close to the thermodynamic limit, but not in the limit. All extensive quantities are, therefore, proportional to the extent of the system. We show that the statistical fluctuations depend on the ensemble used. The thermodynamic variables describing the ensemble *uniquely* determine the magnitudes of all *allowed* fluctuations in the system. We argue that the number of *independent* fluctuations is strictly equal to $n$, where $n$ is the number of extensive quantities with respect to which the system is open to the surrounding; we will refer to $n$ as *the thermodynamic degree* of the ensemble. One of the results of the fluctuation theory is similar to the uncertainty relation in quantum mechanics. The similarity is probed. We prove five useful theorems. We explain how the general fluctuation theory can be extended to stationary metastable states. We follow the consequences of vanishing susceptibility for Nernst points. In particular, its consequence for stationary metastable states is discussed. We argue that such states *cannot* exist in Nature.[7] We consider monomer number fluctuations. We show that in a special constant volume ensemble (see Sec. VII), it is possible for the total monomer fluctuation $<(\Delta N_m)^2>$ to be zero so that there is no density fluctuation even if the individual monomer numbers fluctuate to give rise to composition fluctuations. In this case, the composition fluctuation *cannot* contribute to $<(\Delta N_m)^2>$. Indeed we will see that $<(\Delta N_m)^2>$ does not contain any explicit composition fluctuation. Thus, neither of its parts refers to the composition fluctuation. The intensity, on the other hand, can contain composition fluctuation. Thus, the presence of the K- and µ-parts in $I(0|b_1,b_2)$ and $<(\Delta N_\mathrm{m})^2>$ says nothing about either of the two claims above. To underscore this observation, we show that, in two of the other ensembles that we study here, $<(\Delta N_\mathrm{m})^2>$ for athermal symmetric blends contains only the K-part, even though there are composition fluctuation and cross-fluctuation in the system.

The layout of the paper is as follows. We consider the general framework in the following section. We introduce various ensembles in Sect. III. In Sec. IV, we present the basis of the fluctuation theory following Ref. 1 and provide two simple examples to illustrate the formalism. We present five formal results in Sec. V in the form of Theorems 1-4 and a corollary valid for general fluctuations that prove extremely useful. We investigate Nernst point[2] at which fluctuation in some extensive quantity vanishes in Sect. VI and prove a generalized Nernst Theorem. This point plays an important role in Sec. X, and has been discussed in detail in Ref. 5. In the following three sections, we consider fluctuations in three different ensembles, and explicitly demonstrate that there is in general cross-correlation between density and composition fluctuations, making them *statistically dependent*. In Sec. X, we focus on the density and composition fluctuations formally and show that it is incorrect to claim that the K- and the μ- terms in the intensity and the total monomer number fluctuation refer to density and composition fluctuations, respectively in the grand canonical ensemble. We demonstrate that $\Delta N_\mathrm{m}$ does not depend explicitly on the composition fluctuation. Thus, $<(\Delta N_\mathrm{m})^2>$ *cannot* contain any composition fluctuation, even if it contains two parts. We show that it is meaningless to talk about density fluctuation at constant composition, except at $T = 0$ (a Nernst point), showing that the two fluctuations are indirectly intertwined. We also extend a well-known relation between number and volume fluctuation valid for a single-component system to a multi-component system. We explicitly consider an athermal symmetric blend, which exemplifies our results. The final section contains a discussion of how to extend our results to arbitrary and different hard-core volumes and a brief summary.

## II. General Framework

The discussion of the fluctuation requires focusing on a given *region* $\mathcal{R}$ of the thermodynamic system. The extent or the size of the region $\mathcal{R}$ is *fixed* by *fixing* the value of one of the $d$ maximum allowed thermodynamic extensive quantities $X$. One usually takes this to be the volume $V$. However, one can choose any one of the extensive quantities in the system, like the number of particles, the internal energy $E$, etc. One can take more than one extensive quantity to characterize $\mathcal{R}$. However, we need *at least* one extensive quantity to specify the size of the region $\mathcal{R}$. Because of this, we obtain a very important bound on the thermodynamic degree of an ensemble: $n \leq d - 1$.

In general, the entropy $S(\mathrm{X})$ is a function of $d$ extensive quantities $X_1, X_2, X_3,\ldots, X_d$, to be collectively denoted by the set X:

$$S(\mathrm{X}) \equiv S(X_1, X_2, X_3,\ldots, X_d).$$

We define the entropy in the units of the Boltzmann constant $k_\mathrm{B}$ so that $S$ is *dimensionless*. All of the extensive quantities are *fixed* in the microcanonical ensemble. Because of this, none of the $d$ quantities are allowed to *statistically* fluctuate in this ensemble: We say that the system is *isolated* from the surrounding. In other words, it forms a completely *closed* system ($n = 0$). The entropy $S(\mathrm{X})$ is the "free energy," i.e. the thermodynamic potential in the microcanonical ensemble. In other ensembles, the situation is different. There are usually $n < d$ extensive quantities, to be labeled $X_1$, $X_2$, $X_3$,…, $X_n$, that are *summed* over in the partition function, so that they are allowed to *statistically fluctuate*. We will denote the set $\{X_k\}$ of fluctuating extensive quantities by $X_F$; the number of elements in $X_F$ represents the thermodynamic degree of the ensemble. This leaves behind $f \equiv d-n$ extensive quantities that remain *fixed*. As said above, we must have $d \geq f \geq 1$. The system is said to be *open* to the surrounding with respect to these *fluctuating* extensive quantities. All of the $X_k \in X_F$ fluctuate simultaneously, their fluctuations being controlled by the physical properties of the surrounding. Corresponding to each fluctuating $X_k$, there is a conjugate field $Y_k$, imposed by the surrounding on the system. It has the property that the product $X_i Y_i$ is *dimensionless*, like the entropy. The conjugate field is a property of the surrounding system. For the energy $E$, the volume $V$, the number of particles $N_p$,…, the conjugate fields are $(-\beta)$, $(-\beta P)$, $\beta \mu_p$, …, respectively, where $\beta \equiv 1/T$, $T$ being measured in the units of the Boltzmann constant $k_B$, $P$ is the pressure, and $\mu_p$ is the chemical potential per particle. In equilibrium, these fields are also the fields in the system, for which

$$Y_i \equiv y_i / T \equiv -(\partial S / \partial X_i)_{X'_F}, \tag{3}$$

where $X'_F$ denotes the rest of the set X besides $X_i$ that must be kept fixed during differentiation, and we have introduced a new related field $y_i$ by taking out *T*. However, since the system is open to the surrounding as far as $X_k \in X_F$ are concerned, there are fluctuations not only in $X_k$, but also in their conjugate fields $Y_k$, even in equilibrium. We will denote the set of conjugate fields $Y_k$, $k \leq n$, by $Y_F$. In the following, we are interested in statistical fluctuations in $X_k \in X_F$ and $Y_k \in Y_F$. The remaining $f \geq 1$ extensive quantities $X_{n+1}$, $X_{n+2}$, $X_{n+3}$,….., $X_d$, form a set of *fixed*, i.e. non-fluctuating extensive quantities, and we will use $X_{NF}$ to collectively denote them as a set. No fluctuation can occur in $X_k \in X_{NF}$. For example, in the canonical ensemble, the internal energy $E$ is allowed to fluctuate with its fluctuation controlled by the inverse negative temperature $(-\beta)$. But there are also fluctuations[1] in *T*; see below also. The remaining extensive quantities form the set $X_{NF}$ in this ensemble.

There is another *important* reason for specifying $\mathcal{R}$ by some extensive quantity or quantities. This has to do with the statistical mechanical description of the problem using a partition function. We introduce a sequence of the partition functions $\{Z_N\}$, indexed by some extensive quantity $N$ so that the *thermodynamic limit* can be properly taken, as $N$ diverges. We construct a sequence of *intensive* "free energy" $\{\omega_N\}$, where

$$\omega_N \equiv (1/N) \ln Z_N.$$

The sequence $\{\omega_N\}$ is expected to converge to some limit $\omega$ as the extent of $\mathcal{R}$ diverges. In this limiting process, an extensive quantity $Q_N$ is expected to possess a limiting density, which is the limit of the sequence $\{Q_N / N\}$. In the case when more than one extensive quantity are kept fixed, we choose arbitrarily one of them to index the partition function.

Consider a single-component system. Here, the set X contains $E, V$, and $N_p$. (We do not consider the system to be, for example, magnetized or electrically polarized.) From the discussion above, it should be clear that one could define a constant $T$, constant $P$ ensemble by allowing fluctuations in $E$ and $V$, so that $X_F \in E, V$. The corresponding thermodynamic potential for this ensemble is the Gibbs free energy. We *cannot* allow $N_p$ to be part of $X_F$, as it will be required to index the partition function to obtain the thermodynamic limit; see for example, Ref. 35. The Gibbs free energy is proportional to $N_p$ for a finite but macroscopically large system.

By restricting the allowed states in the definition of the partition function, we can use the same statistical mechanical framework to describe *metastable states* in a system, which do not depend on time. Such time-independent metastable states are supposed to be the long time limits of physically observed metastable states. In this sense, we are considering stationary limits of metastable states, which we will call here *stationary metastable states*. Since such metastable states no longer depend on time of observation $t$, the partition function formulation is perfectly suited for them. The restriction or restrictions are imposed so that the equilibrium states are forbidden in the partition function. As shown in Ref. 7, the partition function formulation, which represents a sum of only non-negative terms, ensures not only that the stability requirements of non-negative susceptibilities like the heat capacity, the compressibility, etc. are obeyed by stationary metastable states, but also that the restricted free energy $\omega$ must achieve its *maximum* value for the stationary metastable states. Thus, stationary metastable states behave similar to equilibrium states except that the equilibrium states have their free energy $\omega$ *higher* than that of stationary metastable states.

## III. Ensembles

We will demonstrate in this review by several examples that the fluctuations depend on the statistical ensemble used for the calculation; see also Refs. 26, and 28. Therefore, one must carefully choose the proper ensemble that suits the experimental setup. We will consider a binary mixture below for ease of discussion, which is easily extended to an *r*-component mixture. A binary mixture of two polymeric species 1 and 2 contains $d = 2+r = 4$ extensive quantities $V, E, N_{m1}$ and $N_{m2}$. We will assume that each monomer of the *j*-th species has a *constant* volume $v_j$, which can be identified as its *hard-core volume*. In the following, we take $v_1 = v_2 = v_0$. In the last section, we will extend the results to the case when $v_1$ and $v_2$ are different. The difference

$$V_0 \equiv V - (N_{m1} + N_{m2})v_0, \quad (N_0 \equiv V_0 / v_0) \tag{4}$$

will be called the *free volume* in the following. In terms of $V_0$, we have introduced $N_0$, which can be identified with the *number* of *voids* in a lattice model of the system; otherwise, it is a dimensionless number as a measure of the free volume. From now on, we will set $v_0=1$ for simplicity of formulation. This does not affect the results. Our discussion is general enough to be applicable to both the continuum and lattice models.

One of the simplest, but the least convenient of the ensembles is the *microcanonical* ensemble in which all the $d = 4$ extensive quantities are kept *fixed*; thus, $n = 0$. The ensemble does not allow for *any* fluctuation and does not have to be considered here. The next level of ensembles corresponds to $n = 1$; here, three of the extensive quantities are kept fixed. If we keep $V, N_{m1}$ and $N_{m2}$ fixed, but sum over $E$, we obtain the customary *canonical* ensemble ($X_F = \{E\}$, $X_{NF} = \{V, N_{m1}, N_{m2}\}$). The system is now open with respect to energy exchange with the surrounding; the energy fluctuates and is controlled by its conjugate *field* ($-\beta$). This ensemble does not allow for any fluctuations in the number of monomers. If we open the system so as to exchange monomers of only one species by summing over it, and keep the remaining three quantities fixed, it allows for fluctuations in only the number of monomers of this species. If we sum over the numbers of monomers of both species, we obtain the *grand canonical ensemble*, which allows for fluctuations in the energy and both monomer numbers in a constant volume. Of course, we can allow for fluctuations in $N_{m1}$ and $N_{m2}$, and keep $V$ and $E$ fixed. However, no experiments are done at constant $E$; rather, they are performed at constant $T$. Therefore, the useful ensemble must not keep $E$ fixed. At this stage, we *cannot* allow $V$ to fluctuate in the grand canonical ensemble.

Three ensembles labeled A, B, and C, have been introduced in Ref. 26, in which $E$ is *not* fixed. Despite this, the ensembles allow $N_{m1}$ and $N_{m2}$ to fluctuate. The A-ensemble ($n = 2$) contains two fixed extensive quantities $V$ and $V_0$, and the system is open with respect to exchanges of $E$ and $N_{m1}$. The conjugate fields are ($-\beta$) and $\beta\bar{\mu}$, $\bar{\mu} \equiv \mu_{m1} - \mu_{m2}$, respectively and the partition function is given by

$$Z_A(T, \bar{\mu} | V, V_0) \equiv \sum_{E, N_{m1}} \exp[S - \beta E + \beta\bar{\mu} N_{m1}], \tag{5}$$

where the summation over $E$ must be interpreted as an integration over $E$ for a continuum model. Moreover, $S(E, N_{m1} | V, V_0)$ is the entropy of the system and $N_{m2}$ is given by $N_{m2} = V - V_0 - N_{m1}$; see Eq. (4). Here, $Y_F = \{T, \bar{\mu}\}$, $X_F = \{E, N_{m1}\}$, and $X_{NF} = \{V, V_0\}$. Since there is no fluctuation in $V$ and $V_0$, we have

$$\Delta N_{m2} = -\Delta N_{m1}, \quad \Delta N_m \equiv 0. \tag{6}$$

Thus, both monomer numbers fluctuate, though in a *strongly correlated* manner. Fixing $V_0$, and $V$ is equivalent to keeping the total monomer number $N_m \equiv N_{m1} + N_{m2}$ fixed.

The B-ensemble ($n = 3$) is obtained from the A-ensemble by summing over $V$ by the use of the associated conjugate field ($-\beta P$). Only the free volume is kept fixed. The partition function is now

$$Z_B(T, \bar{\mu}, P | V_0) \equiv \sum_{E, N_{m1}, V} \exp[S - \beta(E - \bar{\mu} N_{m1} + PV)], \tag{7}$$

where the entropy $S(E, N_{m1}, V | V_0) \equiv S(E, N_{m1} | V, V_0)$. Here, $Y_F = \{T, \bar{\mu}, P\}$, $X_F = \{E, N_{m1}, V\}$, and $X_{NF} = \{V_0\}$. Again, the summation over $V$ must be interpreted as a volume integral for a continuum model. It is obvious that, compare with Eq. (6),

$$\Delta N_{m2} = \Delta V - \Delta N_{m1}, \quad \Delta N_m = \Delta V. \tag{8}$$

However, the most convenient ensemble for the calculation of fluctuations in $N_{m1}$ and $N_{m2}$ is the constant volume ensemble, called the C-ensemble ($n = 2$) by us.[26] It is the standard *grand canonical ensemble* in which the volume $V$, the temperature $T$ and the two chemical potentials $\mu_{m1}$ and $\mu_{m2}$ are kept fixed. The corresponding partition function is given by

$$Z_C(T, \mu_{m1}, \mu_{m2}|V) \equiv \sum_{E, N_{m1}, N_{m2}} \exp[S - \beta E + \beta(\mu_{m1} N_{m1} + \mu_{m2} N_{m2})], \quad (9)$$

where $S(E, N_{m1}, N_{m2}|V) \equiv S(E, N_{m1}|V, V_0)$, with $V_0$ determined by Eq. (4). Here, $Y_F = \{T, \mu_{m1}, \mu_{m2}\}$, $X_F = \{E, N_{m1}, N_{m2}\}$, and $X_{NF} = \{V\}$.

For the A and C ensembles, $V$ is used as the index for the partition function and is allowed to diverge to infinity, keeping all the densities fixed and finite for the thermodynamic limit. For the B ensemble, this role is played by $V_0$. The extension of all three ensembles to an $r$-component mixture is trivial.

## IV. Fluctuation Theory

We will closely follow Landau, Lifshitz and Pitaevskii[1] for our calculation of statistical fluctuations not only in $N_{mj}$, but also in other quantities in an ensemble of thermodynamic degree $n$. The theory assumes *small* fluctuations, but we will demonstrate by two examples at the end of the section that the results are valid generally regardless of the magnitude of the fluctuations. This approach is based on the original ideas of Einstein.[36] The probability of fluctuation is controlled by a quantity $R$ given by

$$R = 2(\Delta S + \sum_{1 \leq k \leq n} Y_k \Delta X_k),$$

where the sum is over *all* of the $n$ extensive quantities $X_k$ with respect to which the system is open. By introducing $X_0 \equiv S$ and $Y_0 \equiv 1$, ($y_0 = T$), we can include the first term under the summation by extending it over $k$ from 0 to $n$. [It should be stressed that $X_0, Y_0$ are not related via Eq. (3).] Expanding the entropy fluctuation to second order in terms of $\Delta X_k$, $1 \leq k \leq n$ and recalling Eq. (3), we can rewrite $R$ as a sum over the products $\Delta X_k \Delta Y_k$:

$$R = -\sum_{1 \leq k \leq n} \Delta X_k \Delta Y_k. \quad (10)$$

We should remark that our formalism is slightly different from that in Ref. 1, which is obtained by expanding the internal energy $X_1 \equiv E$ in terms of $S$ and the remaining set $X'$. The result is that $R$ is now given by[1]

$$RT = -(\Delta T \Delta S + \sum_{2 \leq k \leq n} \Delta y_k \Delta X_k), \quad (11)$$

where $y_i$ is introduced in Eq. (3). The first term in Eq. (11) can also be written as $\Delta y_0 \Delta X_0$, and can be included in the sum.

The probability distribution of fluctuations is given by

$$W(\{\Delta X_k \mid X_k \in X_F\}) = W_0 \exp(R/2),$$

where $W_0$ is a normalization constant. According to Ref. 1, we can express $R$ in terms of any $n$ independent variables, *not* all of them have to be extensive. We will take them to be such that no two of them form a conjugate pair [$Y_k$ and $X_k$ for Eq. (10) or $y_k$ and $X_k$ for Eq. (11)]. The results for fluctuations will be *independent* of this choice. This can easily be checked by simple calculation.

As argued elsewhere,[1,5] the above form of $W$ immediately gives rise to the following "uncertainty-relation" between $X_k$ and $Y_l$:

$$< \Delta X_k \Delta Y_l > = \delta_{kl}. \qquad (12)$$

Accordingly, a *field* $Y_k$ couples only with its own conjugate extensive quantity $X_k$, but not with other extensive quantities $X_l$, $k \neq l$, which follows from the definition of the Kronecker delta in Eq. (12). *This result is true in every ensemble.*

Since the entropy is expanded to second order in terms of $\Delta X_k$, $1 \leq k \leq n$, in obtaining Eq. (10), we find from Eq. (3) that $Y_i$ is a linear combination of $\Delta X_k$. Let us express this relation as follows:

$$Y_i = \sum_k \rho_{ik} \Delta X_k, \qquad (13)$$

where $\rho_{ik}$ forms a symmetric matrix. $\boldsymbol{\rho}$ Substituting this in Eq. (12), and observing that $< \Delta X_k > \equiv 0$, we conclude that

$$< \Delta X_k \Delta X_l > = (\boldsymbol{\rho}^{-1})_{kl}, \qquad (14)$$

where $\boldsymbol{\rho}^{-1}$ is the inverse of the matrix $\boldsymbol{\rho}$. Finally, we have

$$< \Delta Y_k \Delta Y_l > = \rho_{kl}. \qquad (15)$$

In terms of $y_l$, the uncertainty relation reduces to

$$< \Delta X_k \Delta y_l > = T \delta_{kl}. \qquad (16a)$$

In terms of the density $x_k \equiv X_k / V$, the same relation becomes

$$< \Delta x_k \Delta y_l > = (T/V) \delta_{kl}. \qquad (16b)$$

Thus, we conclude that fluctuations in the density $x_k$ and its conjugate field $y_k$ are *suppressed* as $T \to 0$, and/or $V \to \infty$, just as the quantum fluctuations are suppressed in the classical limit Planck's constant $\hbar \to 0$.[5] In other words, $T/V$ plays the role of the Planck's constant $\hbar$.

We consider two simple examples before proceeding.

**Example 1.** Consider $n=1$, with $X_F = \{E\}$. This is the standard canonical ensemble. Here, $R = \Delta \beta \Delta E$. The set $X_{NF}$ contains all the remaining fixed (non-fluctuating) extensive quantities (except $E = X_1$). Expressing $\Delta E = (\partial E / \partial \beta)_{X_{NF}} \Delta \beta$, we have $R = (\partial E / \partial \beta)_{X_{NF}} (\Delta \beta)^2$. Thus, $< (\Delta \beta)^2 > = -1/(\partial E / \partial \beta)_{X_{NF}}$. We can also express $\Delta \beta = (\partial \beta / \partial E)_{X_{NF}} \Delta E$ to get $< (\Delta E)^2 > = -(\partial E / \partial \beta)_{X_{NF}}$. Thus, we finally get

$$< (\Delta T)^2 >_{can} = T^2 / C_{X_{NF}}, \quad < (\Delta E)^2 >_{can} = T^2 C_{X_{NF}}, \quad < \Delta \beta \Delta E >_{can} = -1, \qquad (17)$$

where $C_{X_{NF}}$ denotes the heat capacity $(\partial E / \partial T)_{X_{NF}}$ at constant $X_{NF}$, and the subscript "can" stands for the canonical ensemble. At a critical point, $C_{X_{NF}}$ diverges and the fluctuation in $T$ becomes zero, while the fluctuation in $E$ diverges. However, $C_{X_{NF}} = 0$ at $T = 0$ (Nernst point) and the fluctuation in $T$ diverges; the fluctuation in $E$, however, vanishes and $E$ is frozen at its minimum possible value. From Eq. (3), we note that field $y$ corresponding to $E$ is $(-1)$, since the corresponding $Y = -\beta$. Therefore, the fluctuation in $y$ also vanishes as we discussed above: $T \to 0$ suppresses fluctuations.

Since the derivatives always have to be taken at fixed $X_{NF}$, we will usually suppress it as part of various derivatives in the following.

**Example 2. (D-ensemble).** Consider $n=2$, with $X_F = \{E, X_2 \equiv V\}$, and $Y_F = \{-\beta, -\beta P\}$. The set $X_{NF} = \{N_{m1}, N_{m2}\}$ will not be exhibited in the derivatives except in the final result. We will name this ensemble the D-ensemble. [If $X_{NF}$ had contained only one extensive quantity $N_p$, the number of particles, then the resulting ensemble would the standard $T - P - N_p$ ensemble.] We have

$$R = \Delta\beta\Delta E + \Delta(\beta P)\Delta V.$$

It is convenient to expand $\Delta E$ and $\Delta(\beta P)$ in terms of $\Delta T$ and $\Delta V$:

$$\begin{aligned}\Delta E &= (\partial E/\partial T)_V \Delta T + (\partial E/\partial V)_T \Delta V, \\ \Delta \beta P &= (\partial \beta P/\partial T)_V \Delta T + (\partial \beta P/\partial V)_T \Delta V.\end{aligned} \quad (18)$$

Thus,

$$R = -(1/T^2)(\partial E/\partial T)_V (\Delta T)^2 + (1/T)(\partial P/\partial V)_T (\Delta V)^2,$$

so that,

$$\begin{aligned}<(\Delta T)^2>_D &= T^2/C_{V,X_{NF}}, \quad <(\Delta V)^2>_D = TVK_{T,X_{NF}}, \\ <\Delta T \Delta V>_D &= 0,\end{aligned} \quad (19)$$

where we have now explicitly shown $X_{NF}$ as part of the derivatives. Here, $K_{T,X_{NF}}$ denotes the compressibility $-(1/V)(\partial V/\partial P)_{T,X_{NF}}$ and the last relation is in accordance with Eq. (12). We square $\Delta E$, and use Eq. (19) to obtain

$$<(\Delta E)^2>_D = T^2 C_{V,X_{NF}} + TVK_{T,X_{NF}}[T(\partial P/\partial T)_{V,X_{NF}} - P]^2, \quad (20)$$

where the quantity in the square bracket is $(\partial E/\partial V)_{T,X_{NF}}$. It should be noted that the set $\{V\} \cup X_{NF}$ in Eq. (20) is the same as the set $X_{NF}$ in Eq. (12). The above equation clearly shows that the two fluctuations in the internal energy $E$ in Eqs. (17), and (20) are different because of the last term in Eq. (20), which originates from the fluctuations in $V$. However, the temperature fluctuation is the same in both ensembles. These results are examples of the two theorems proved in the next section..

The reason why the energy fluctuation is different in the two cases is not hard to understand. For $n=2$, there is an additional contribution to $\Delta E$ from the volume fluctuation $\Delta V$, as seen from Eqs. (18), and (20). The fluctuating volume contribution is determined by the compressibility, see Eq. (19), and the square of $(\partial E/\partial V)_{T,X_{NF}}$. The first term in Eq. (20) is precisely the same as the energy fluctuation in Eq. (17). Because of the absence of cross-correlation between the temperature and volume fluctuations, see Eq. (19), there are only two separate contributions in the energy fluctuations. Since the additional contribution from the volume fluctuation *cannot* be negative, we have an important result

$$<(\Delta E)^2>_D \geq <(\Delta E)^2>_{can}. \quad (21)$$

Eq. (21) is a special case of a more general result (Theorem 2 below); see Eq. (24a).

Consider Eq. (18) for $E$ and $P$; each one has two different fluctuation contributions. This is a general feature of the fluctuation theory, which we state below as Theorem 1. The number of fluctuating terms including cross-correlation terms in the square of the fluctuation of a given quantity, or the product of two fluctuating quantities, increases as $n$ increases, i.e., as the number of fluctuating quantities increases. Usually, each term will contribute to the average fluctuation. Thus, it is clear that the value of the fluctuation depends strongly on the ensemble.

It is easy to see that the energy fluctuation in Eq. (20) can be rewritten as

$$<(\Delta E)^2>_D = -(\partial E/\partial \beta)_{\beta P, X_{NF}} = T^2 C_{\beta P, X_{NF}}.$$

To see this, we write the derivative $(\partial E/\partial \beta)_{\beta P, X_{NF}}$ as a Jacobian and manipulate it:

$$\partial(E,\beta P)/\partial(\beta,\beta P) = [\partial(E,\beta P)/\partial(T,V)]/[\partial(\beta,\beta P)/\partial(T,V)].$$

The denominator is simply $-(\partial P/\partial V)_{\beta P, X_{NF}}/T^3$. The numerator can be expanded and manipulated to give the right hand side of Eq. (20) with a minus sign. The result can also be obtained by using the partition function, see Eq. (7) but without a sum over $N_{m1}$. Thus, the effect of *fluctuating volume* is to replace the *constant-volume* derivative with a *constant conjugate field βP* derivative. In other words, the derivative is always calculated at *fixed set of extensive quantities* and *fields* belonging to the set $X_{NF} \cup Y_F$. This is an example of a general result, which is presented below as Theorem 2. Moreover, from Eq. (21), we observe that $C_{P,X_{NF}} \geq C_{V,X_{NF}}$. We note that at absolute zero (Nernst point), the heat capacities $C_{V,X_{NF}}$ and $C_{P,X_{NF}}$ ($P \neq 0$) vanish simultaneously. Thus, the energy fluctuation goes to zero, but the temperature fluctuation diverges.[5]

We can now calculate $<\Delta T \Delta E>$ by multiplying Eq. (18) by $\Delta T$ and using (19). We note that $<\Delta T \Delta E> = T^2$, as expected; see Eq. (12). We can also calculate the fluctuation in $P$ and other cross fluctuations. We expand $\Delta P$ in terms of $\Delta T$ and $\Delta V$, as in Eq. (18). We square it or multiplying it by $\Delta V$ and $\Delta T$ separately, and use Eq. (19) to finally obtain

$$<(\Delta P)^2>_D = T/VK_{S,X_{NF}}, \quad <\Delta P \Delta V>_D = -T,$$
$$<\Delta P \Delta T>_D = T^2(\partial P/\partial T)_{V,X_{NF}}/C_{V,X_{NF}} \quad (22)$$

This shows that the pressure fluctuation is correlated with the volume and the temperature fluctuations in the D-, i.e., the $T-P-X_{NF}$-ensemble. We observe that the pressure fluctuation satisfies the general result in Eq. (12), where $Y = (-\beta P)$ and $X = V$. The set $X'$ contains $S$ and $X_{NF}$.

**General Validity of the theory.** Even though the fluctuation theory presented here assumes small fluctuations, the results obtained above are valid in general. The volume fluctuations–compressibility relation $<(\Delta V)^2>_D = TVK_{T,X_{NF}}$ in Eq. (19) is a general relation valid beyond the validity of the fluctuation theory. Indeed, it is valid everywhere including the region near a critical point. This can be easily established by considering a constant pressure ensemble like the D-ensemble. The partition function for the D-ensemble [compare with Eq.(7)] is given by

$$Z_{X_{NF}} \equiv \sum_{E,V} \exp[S(E,V) - \beta E - \beta PV]. \qquad (23a)$$

Differentiating $\ln Z_{X_{NF}}$ with respect to $(-\beta P)$ yields the average value $\overline{V}$ of the volume. Differentiating $\overline{V}$ with respect to $(-\beta P)$ immediately leads to the above relation in Eq. (19). Thus, the results of the fluctuation theory have more generality than their actual derivation. Similarly, the fluctuation $<(\Delta P)^2>$ can be obtained by considering a constant volume ensemble; see Hill,[2] who has calculated it in the canonical ensemble. We find that

$$<(\Delta P)^2>_{can} = T[(\partial P/\partial V)_{T,X_{NF}} - (\partial P/\partial V)_{S,X_{NF}}] = T^2(\partial P/\partial T)^2_{X_{NF}}/C_{V,X_{NF}}. \quad (23b)$$

The result differs from that in Ref. 2, Eq. (19.8), in that the second $(\partial P/\partial V)$ derivative in the first equation is at constant $S$; we omit the arguments leading to it. Setting $\Delta V = 0$ in Eq. (18), we can also calculate $<(\Delta P)^2>_{can}$ using the fluctuation theory. We immediately find the above result in Eq. (23b). Again, the fluctuation theory and the direct fluctuation calculation from the partition function yield the same result. This justifies our generality claim. The contribution when $\Delta V \neq 0$ appears in Eq. (22).

## V.  Some Useful Theorems

**Theorem 1**: A fluctuating extensive or field quantity $\Delta Q$ generally has $n$ statistically independent fluctuating contributions.

**Proof**. As noted in Sect. IV, $R$ is expanded as a bilinear combination in terms of $n$ fluctuations $\Delta X_k$, $X_k \in X_F$. It is given by

$$R = \sum_{k,l=1}^{n} \rho_{kl} \Delta X_k \Delta X_l,$$

where $\rho_{kl} \equiv (\partial^2 S/\partial X_k X_l)$ evaluated at $\Delta X_j=0$, $X_j \in X_F$, are symmetric coefficients.[1] The corresponding symmetric matrix $\boldsymbol{\rho}$ can be diagonalized and $R$ can be expressed in terms of $n$ independent fluctuations $\Delta X'_k$, which are expressible as linear combinations of $\Delta X_k$. In terms of $\Delta X'_k$, there are no cross-fluctuations.[1] Thus, $\Delta X'_k$ form a statistically independent set. We refer the reader to Ref. 1 for details. Each fluctuation, when expressed as a linear combination of $\Delta X'_k$, will have $n$ statistically independent contributions. Q.E.D.

**Theorem 2**: The fluctuations in an extensive quantity $X \neq X_k$, $X_k \in X_F$, in the two ensembles determined by the fixed set $X_{NF} \cup \{X\} \cup Y_F$ and $X_{NF} \cup \{Y\} \cup Y_F$ are given by $(\partial X_k/\partial Y_k)_{X_{NF},Y'_F,X}$ and $(\partial X_k/\partial Y_k)_{X_{NF},Y'_F,Y}$ respectively. Here, $Y$ and $Y_k$ are fields conjugate to $X$ and $X_k$, respectively, and $Y'_F$ is $Y_F$ without $Y_k$. Furthermore,

$$<(\Delta X_k)^2>_{X_{NF},Y'_F,Y} \geq <(\Delta X_k)^2>_{X_{NF},Y'_F,X} \quad (24a)$$

**Proof**: The proof of the theorem is straightforward. Let $Z(Y_k, Y'_F | X_{NF}, X)$ be the partition function for the fixed set $X_{NF} \cup \{X\}$ in which the field $Y_k$ controls the fluctuation of the extensive quantity $X_k$. The fluctuation is given by

$$< (\Delta X_k)^2 >_{X_{NF}, Y_F, X} = (\partial X_k / \partial Y_k)_{X_{NF}, Y'_F, X}.$$

The second partition function is given by

$$Z(Y, Y_k, Y'_F | X_{NF}) = \sum_X Z(Y_k, Y'_F | X_F, X) \exp(XY).$$

Now, the fluctuation in $X_k$ again requires differentiating, but this time keeping the new field $Y$ fixed, with the result

$$< (\Delta X_k)^2 >_{X_{NF}, Y_F, Y} = (\partial X_k / \partial Y_k)_{X_{NF}, Y'_F, Y},$$

which proves the first part of the theorem.

To prove the last part of the theorem, we expand $\Delta X_k$ in terms of $\Delta Y_l$, $Y_l \in Y_F$, and $\Delta X$. The contributions from $Y_l \in Y_F$ give the right-hand side of Eq. (24a). From Eq. (12), we know that $\Delta X$ and $\Delta Y_l$ are statistically independent. Hence, the new contribution to $(\Delta X_k)^2$ comes from the square of the $\Delta X$ term and must be non-negative. This proves the last part of the theorem. Q.E.D.

Consequently, we see that the fluctuation in an extensive quantity is always given by a derivative at constant set $\{Y\} \cup Y'_F$ of fields. Of course, the derivative is also carried out additionally at constant $X_{NF}$. A similar theorem below also holds for fluctuation in the field $Y_k$, except that the quantities held constant are the extensive quantities.

**Theorem 3**: The fluctuation $< (\Delta y_i)^2 >$ in a field $y_i$ is given by $T / (\partial X_i / \partial y_i)_{X'}$, where X′ is defined below.

**Proof**: The proof is simplified by considering the form of $R$ in Eq. (11). Let X={$S, X_2, X_3, \ldots, X_d$} be the set of all extensive quantities including $S$ but not $X_1 = E$, and let X′ denote the remaining set X besides $X_i$. Recall that $X_0 \equiv S$, and $y_0 \equiv T$. We expand all fluctuations in Eq. (11) in terms of fluctuations $\Delta X_k$ ($k \neq i, 2 \leq k \leq n$) and $\Delta y_i$; $i=0$ is allowed in this proof. The coefficient of $(\Delta y_i)^2$ in $R$ is obviously $-(\partial X_i / \partial y_i)_{X'}$. According to Eq. (12), there is no correlation between $\Delta y_i$ and quantities in the set X′. Therefore, the fluctuation is given by

$$< (\Delta y_i)^2 > = T / (\partial X_i / \partial y_i)_{X'}, \quad (24b)$$

and remains the same in all ensembles. This proves our theorem. Q.E.D.

**Theorem 4:** The cross-fluctuation $< \Delta X \Delta X' >_{X_{NF}, Y_F}$ is related to the self-fluctuation $< (\Delta X)^2 >_{X_{NF}, Y_F}$ by

$$< \Delta X \Delta X' >_{X_{NF}, Y_F} = (\partial X' / \partial X)_{X_{NF}, Y'_F} < (\Delta X)^2 >_{X_{NF}, Y_F}, \quad (24c)$$

where $Y'_F$ is $Y_F$ except $Y$, the field conjugate to $X$.

**Proof:** We observe from Theorem 2 that $< (\Delta X)^2 >_{X_{NF}, Y_F} = (\partial X / \partial Y)_{X_{NF}, Y'_F}$. Also, using Theorem 1, we expand $\Delta X'$ in terms of field fluctuations in $Y_k \in Y_F$:

$$\Delta X' = \sum_k (\partial X' / \partial Y_k)_{X_{NF}, Y'_F} \Delta Y_k.$$

Using this expansion in the product $\Delta X \Delta X'$, and observing from Eq. (12) that only $\Delta X \Delta Y$ has non-vanishing cross-fluctuation, we find that $<\Delta X \Delta X'>_{X_{NF},Y_F} = (\partial X'/\partial Y)_{X_{NF},Y'_F}$, and the theorem follows. Q.E.D.

This theorem shows that cross-fluctuations can be expressed in terms of a self-fluctuations and, therefore, not independent. This is in accordance with Theorem 1 in that any extensive or field fluctuation can always be expressed in terms of $n$ self-fluctuations in extensive quantities. However, the last theorem provides an extension of Theorem 1, which we present as a corollary below.

**Corollary:** A fluctuating quantity $\Delta Q$ generally has at most $n$ different fluctuating contributions.

**Proof:** Instead of considering statistically independent fluctuations $\Delta X'_k$ introduced in Theorem 1, we can express $\Delta Q$ in terms of $\Delta X_k$. Let $X'_F \subseteq X_F$ containing $n' \leq n$ extensive quantities $X_k \in X'_F$ and let $Q(X'_F)$ be a function of only $X_k \in X'_F$. When we square $\Delta Q$, we get $(\Delta X_k)^2$ and cross-terms $\Delta X_k \Delta X_l$. The cross-fluctuations can be expressed in terms of self-fluctuations as Theorem 4 shows. Hence, the fluctuation $<(\Delta Q)^2>$ contains only $n'$ independent fluctuations. This proves the corollary. Q.E.D.

For $i=0$, Eq. (24b) gives the fluctuation in $T$:

$$<(\Delta T)^2> = T^2/C_{X'}, \tag{25}$$

where $C_{X'}$ is the heat capacity at constant $X'=\{X_2,X_3,\ldots,X_d\}$. The fluctuation is the same as in Eqs. (17), and (19). Another example of Eq. (24b) occurs in Eq. (22) above.

## VI. Nernst Points

In this section, we pursue the consequences of *vanishing susceptibilities*, a regime that has been overlooked for the most part in statistical physics. The regime has been recently investigated by us,[5] and the present section is based primarily on this work to which we refer the reader for various applications and details, which are omitted here.

As said in Sect. I, the most famous and well-known consequence of *vanishing* specific heat at $T=0$ is known as the Nernst's theorem.[17] It is usually stated as follows: The entropy $S$ of a system at $T=0$ ($\beta \to \infty$) is a *constant*, which may be taken to be zero. Under some mild assumptions,[4] this leads to the unattainability of $T=0$ ($\beta \to \infty$) in a finite number of steps, which is used as an alternative formulation of Nernst's theorem. In a certain sense, $\beta$ seems to be more of a "natural variable" than $T$ in statistical physics,[5] as it forms the boundary of the inverse temperature scale. In this formulation, crossing over to *negative* temperatures, which are physically realizable,[37] corresponds to crossing $\beta=0$.

The unattainability of $\beta \to \infty$ presumably cannot be distinguished from other similar statements regarding the impossibility of attaining infinitely large $P, H$, etc. in a finite number of physical steps. Indeed, as shown in Ref. 5, this unattainability forms the proper generalization of Nernst's theorem. The above limitation is a consequence of an appropriate vanishing susceptibility like the isothermal compressibility, the isothermal magnetic susceptibility, etc.

The physical significance of Eq. (12) is quite interesting[5] in some cases. Consider $n=1$ and let the conjugate pair be $X$ and $Y$ with respect to which the system is open. In general, they both have statistical fluctuations. The cross-fluctuation is always one. At a critical point the fluctuation in $X$ diverges. This implies that the fluctuation in $Y$ *vanishes*, if Eq. (12) is to be satisfied.

At a *Nernst* point[5] at $T > 0$, the fluctuation in $X$ *vanishes*. The uncertainty relation must still hold. Therefore, the fluctuation in $Y$ becomes *infinitely* large. At this point, the value of $X$ is *frozen* at its equilibrium value with no fluctuation allowed in it, as we prove below in the form of the following theorem, the generalized Nernst's theorem.[5]

**Theorem 5 (Generalized Nernst's Theorem):** The extensive quantity $X$ takes its *extremum* value $X_0$, which is a universal constant in the region where the susceptibility $\chi_R \equiv (\partial X / \partial Y)_R$ *vanishes*, regardless of the set of parameters $R$, provided all other susceptibilities remain finite.

To prove the theorem, we consider a simple system with $n = 2$, with $S$ and $V$ as the two extensive quantities, with corresponding $y$'s given by $T$ and $P$, respectively. In Ref. 5, we had taken the extensive quantities to be $S$ and $M$. The proof is identical. We consider the following thermodynamic identities:

$$C_P - C_V = T(\partial V / \partial T)_P^2 / V K_T = T(\partial P / \partial T)_V (\partial V / \partial T)_P$$
$$K_T - K_S = T(\partial V / \partial T)_P^2 / V C_P \; ; \quad C_P / C_V = K_T / K_S. \tag{26}$$

Here, $C$, and $K$ denote the heat capacity and the compressibility, respectively, and the subscripts have the conventional meaning. Let us assume that the isobaric heat capacity $K_T$ vanishes along some curve $\Gamma$ in the $P-T$ plane. Assuming that other susceptibilities like the heat capacities are finite, we conclude that the thermal expansion derivative $(\partial V / \partial T)_P \to 0$, as $K_T \to 0$, such that

$$0 \leq \left|(\partial V / \partial T)_P^2 / K_T\right| < \infty.$$

As $K_T \to 0$, we conclude that $dV = (\partial V / \partial T)_P dT + (\partial V / \partial P)_T dP \to 0$. Thus, $V$ takes a constant value $V_0$ on $\Gamma$: $V$ is "frozen" to have a fixed value $V_0$ on $\Gamma$. Furthermore, since $(\partial V / \partial T)_P \to 0$, and $(\partial V / \partial P)_T \to 0$ on $\Gamma$, $V_0$ represents the extremum value of $V$. This proves our generalized Nernst theorem.[5]

Similar conclusions are arrived at if $C_P$ vanishes along $\Gamma$. In this case, the entropy $S$ takes a constant value $S_0$ on $\Gamma$. In the case when $\Gamma$ passes through $T = 0$, the theorem fails if $(\partial S / \partial T)_P \equiv C_p / T$ does *not* vanish. In this case, we cannot argue that $(\partial V / \partial T)_P \equiv -(\partial S / \partial P)_T \to 0$, and the theorem fails.

It should be stressed that the proof does not require any stability requirement. Thus, it is valid even if stability is not obeyed. Furthermore, the theorem is valid as long as thermodynamic relations in Eq. (26) are observed. Thus, we can anticipate the theorem to be valid in metastable states, provided Eq. (26) is observed. As a consequence, the theorem is certainly valid for stationary metastable states.

The equilibrium value of $X$ at the Nernst point is its *extremum* value. For example, the case when the heat capacity vanishes at absolute zero or above, then the fluctuations in the energy in the canonical ensemble vanishes; see Eq. (17). The value of the energy is *frozen* at its minimum possible value. On the other hand, the temperature fluctuations become *anomalous*. (For this, the heat capacity must vanish faster than $T^2$ if it vanishes at absolute zero.) Because the equilibrium value is an extremum, the response function $(\partial X/\partial Y)_{X_{NF},Y'_F}$ must vanish at the Nernst point; here, $Y'_F$ denotes the set $Y_F$ without *Y*. The Nernst point usually occurs at the maximum possible strength of the conjugate field *Y*. For energy fluctuation, it occurs at the maximum possible value of *β*, i.e., at *T*=0. For volume fluctuation, it occurs at infinite pressure where the compressibility goes to zero. We refer the reader to Ref. 5 for details.

To discuss non-equilibrium states, we need to distinguish between the instantaneous value of a field like the temperature $T$, and its value fixed by the surrounding environment like the heat bath. We denote the values of the fields of the surrounding by a subscript $s$. The anomalous fluctuations in $T$ imply that even though the average value $\bar{T}$ of $T$ is well defined, and equal to the temperature $T_s$ of the surrounding heat bath, it requires an infinite amount of time $t$ of observation for $\bar{T}=T_s$ to be valid.[1,5] Relaxation of the system over a *finite duration* can bring about only *partial equilibrium*, and $\bar{T}$ will in general have no relationship with $T_s$ on $\Gamma$. There is a clear parallel between a Nernst point and a critical point: both require an infinite amount of time for approach to equilibrium. This is in our opinion the physical significance of the unattainability principle. The fluctuating field ceases to be an appropriate variable to uniquely describe the system as we can observe the system only for a finite time. Similarly, as the isothermal compressibility vanishes, which implies that the adiabatic compressibility also vanishes (since $K_T \geq K_S$, and both are non-negative), the fluctuations in the pressure also diverge; see Eq. (22). In This case, the pressure loses its significance as a thermodynamic state variable.

We can rewrite the temperature fluctuation in Eq. (17) in terms of the fluctuations in the inverse temperature $\beta$:

$$<(\Delta T)^2>/T_s^2 \equiv <(\Delta\beta)^2/\beta_s^2> = 1/C_{X_{NF}} \to \infty. \qquad (27)$$

Here, $\beta_s = 1/T_s$. The fluctuations in $T$, and $\beta$ are much larger than their respective average values, as we approach $\Gamma$. Therefore, near $\Gamma$, they lose their physical significance completely, as they *cannot* be precisely defined over a finite duration due to anomalous fluctuations in them. This aspect of the Nersnt theorem is usually not appreciated. The situation should be contrasted with the observed behavior near a critical point when $\chi_R \to \infty$. Here, the intensive field variables have *no* fluctuations; only extensive quantities *X* have anomalous fluctuations.

The above conclusion can also be applied to metastable states like the ideal glass. In ideal glass,[15,16] the heat capacity vanishes over a finite temperature range $0 \leq T < T_K$, where $T_K$ is known as the Kauzmann temperature. Hence, if ideal glass exists in Nature, then its temperature cannot be defined or measured in a finite duration. Indeed, such a glass can be brought in contact with any system in Nature, no matter what its temperature. Since the heat capacity of the glass is zero, no heat can be exchanged and its temperature cannot change. Thus, an ideal glass can be thought to be in "thermal equilibrium" with every physical system, regardless of the temperature of the latter. This is certainly absurd. Thus, we must conclude that it is impossible to postulate the existence of the *stationary limit* of the metastable supercooled liquid. In other words, the metastable supercooled liquid never reaches a stationary limit, and the dynamics of the state can never be overlooked. [7]

## VII. A-ensemble

We now turn to some concrete calculation. We first consider the A-ensemble in this section. This is a simple and convenient ensemble for the calculation of monomer fluctuations. We recall from Eq. (6) that the two fluctuations $\Delta N_{m1}$ and $\Delta N_{m2}$ are trivially related. Hence, only one of them is needed. However, most importantly for this ensemble,

$$<(\Delta N_m)^2>_A \equiv 0,$$

the subscript A indicating the ensemble used. For $R$, we have from Eq. (10)

$$R = -\Delta E \Delta T / T^2 - \Delta N_{m1} \Delta(\overline{\mu}/T), \tag{28}$$

where we have replace $\Delta\beta$ by $(-\Delta T/T^2)$. Expanding $\Delta E$ and $\Delta(\overline{\mu}/T)$ in terms of $\Delta T$ and $\Delta N_{m1}$, we have

$$R = (-1/T^2)(\partial E/\partial T)_{X_{NF}, N_{m1}} (\Delta T)^2 - (1/T)(\partial \overline{\mu}/\partial N_{m1})_{X_{NF}, T} (\Delta N_{m1})^2, \tag{29}$$

with $X_{NF} = \{V, N_m \text{ or } V_0\}$. The coefficient of $\Delta T \Delta N_{m1}$ in $R$ vanishes identically. Therefore, there is *no* correlation between the temperature and the monomer number fluctuations, as expected from Eq. (12). The fluctuation results are:

$$<(\Delta T)^2>_A = T^2/C_{V, N_{m1}, N_{m2}}, \qquad <(\Delta N_{m1})^2>_A = T(\partial \overline{\mu}/\partial N_{m1})^{-1}_{T, V, N_m},$$
$$<\Delta T \Delta N_{m1}>_A \equiv 0. \tag{30a}$$

Even though $X_{NF}$ contains $V$ and $N_m$, we have set $V, N_{m1},$ and $N_{m2}$ fixed in the heat capacity $C$, since $N_{m1}$ is held fixed in the derivative. We have

$$<\Delta N_{m1} \Delta N_{m2}>_A = -<(\Delta N_{m1})^2>_A, \qquad <(\Delta N_{m2})^2>_A = <(\Delta N_{m1})^2>_A. \tag{30b}$$

One can also check easily that the fluctuation correlation between $\Delta N_{m1}$ and $\Delta \beta \overline{\mu}$ obeys the standard relation in Eq. (12). The first of Eq. (30b) is consistent with Theorem 4. Moreover, $I(0|b_1, b_2)$ from Eq. (1) is given by $(b_1 - b_2)^2 <(\Delta N_{m1})^2>_A$ and is non-zero, even though $<(\Delta N_m)^2>_A \equiv 0$. It is easy to understand the above results from the corollary. The degree of the ensemble is $n = 2$. However, the intensity is a function of only one fluctuating quantity $\Delta N_{m1}$, so that $n' = 1$. Therefore, the intensity depends on only one fluctuating quantity. Let us introduce the total and the individual densities $\rho = N_m/V$, $\rho_j = N_{mj}/V$, $j = 1,2$. From Eq. (30b), we find that

$$<\Delta(\rho_1)^2>_A = <\Delta(\rho_2)^2>_A = -<\Delta\rho_1\Delta\rho_2>_A,$$
$$<\Delta(\rho)^2>_A = 0. \tag{31}$$

There is no total density fluctuation in this ensemble. We have only fluctuations in individual densities, which determine not only $I(0|b_1, b_2)$ but also the fluctuation in the composition $x \equiv N_{m1}/N_m$ since $\Delta x = \Delta N_{m1}/N_m = \Delta \rho_1/\rho_1$. The cross-fluctuation is expressible in terms of self-fluctuation in accordance with Theorem 4 and is not zero. The composition fluctuation $<(\Delta x)^2>_A$ determines $I(0|b_1, b_2)$, but $<(\Delta N_m)^2>_A$ is determined by the total density fluctuation in this ensemble, which is zero. The compressibility of the system plays no role.

## VIII. B-ensemble

We now allow the volume to fluctuate by introducing the pressure $P$ in the problem. The set $X_{NF} = \{V_0\}$ now contains only the free volume $V_0$. Therefore, it is not a physically realizable ensemble. Despite this, as said earlier, it is the only possible constant $T$-$P$ ensemble in which both monomer numbers are allowed to fluctuate. We first observe from Eq. (8) that

$$<(\Delta V)^2>_B \; = \; <(\Delta N_m)^2>_B . \tag{32}$$

The fluctuation in the total number of particles is exactly the same as the volume fluctuation and is, therefore, related to the compressibility of the system [see Eq. (36)], which is different from that in Eq. (19). It is, therefore, clear that the fluctuations in this ensemble are different from the A-ensemble. The two monomer fluctuations are related by $\Delta N_{m2} = \Delta V - \Delta N_{m1}$, see Eq. (8).

We note that there is an additional contribution from $\Delta(\beta P)\Delta V$ to $R$ in Eq. (20) for the A-ensemble. We expand each contribution in $R$ in terms of $\Delta\beta, \Delta V$ and $\Delta N_{m1}$. Using the following expansions, in each of which all derivatives also have $X_{NF} = \{V_0\}$ held constant, which we do not show explicitly,

$$\Delta E \; = \; (\partial E / \partial \beta)_{V,N_{m1}} \Delta\beta \; + \; (\partial E / \partial V)_{T,N_{m1}} \Delta V \; + (\partial E / \partial N_{m1})_{V,T} \Delta N_{m1} ,$$
$$\Delta \beta P \; = \; (\partial \beta P / \partial \beta)_{V,N_{m1}} \Delta\beta \; + \; \beta(\partial P / \partial V)_{T,N_{m1}} \Delta V \; + \; \beta(\partial P / \partial N_{m1})_{V,T} \Delta N_{m1} , \tag{33}$$
$$\Delta \beta \overline{\mu} \; = \; (\partial \beta \overline{\mu} / \partial \beta)_{V,N_{m1}} \Delta\beta \; + \; \beta(\partial \overline{\mu} / \partial V)_{T,N_{m1}} \Delta V \; + \; \beta(\partial \overline{\mu} / \partial N_{m1})_{P,T} \Delta N_{m1} ,$$

in $R$ to express it in terms of $\Delta T, \Delta V$ and $\Delta N_{m1}$, we eventually have

$$R = (\partial E / \partial \beta)_{V,N_{m1}} (\Delta\beta)^2 + \beta(\partial P / \partial V)_{T,N_{m1}} (\Delta V)^2 - \beta(\partial \overline{\mu} / \partial N_{m1})_{V,T} (\Delta N_{m1})^2$$
$$+ \beta(\partial P / \partial N_{m1})_{T,N_{m1}} \Delta V \Delta N_{m1}. \tag{34}$$

We immediately notice that the temperature fluctuations do not couple with any other fluctuations in extensive quantities, as it must be due to Eq. (12). Since constant $V$ and $N_{m1}$ along with constant $V_0$ imply constant $N_{m2}$, the temperature fluctuation here is the same as in the A-ensemble; see Eq.(30a):

$$<(\Delta T)^2>_B \; = \; <(\Delta T)^2>_A , \tag{35}$$

and is in accordance with Theorem 3 for field variables, and Eq. (25). However, there is coupling of volume and monomer fluctuations. This means that these fluctuations will be different from the A-ensemble. We need to consider the 2X2 matrix formed by the coefficients of the remaining fluctuations in Eq. (34), which is

$$\beta \begin{bmatrix} (\partial P / \partial V) & (\partial P / \partial N_{m1}) \\ (\partial P / \partial N_{m1}) & -(\partial \overline{\mu} / \partial N_{m1}) \end{bmatrix},$$

where we have suppressed the quantities that are held fixed for simplicity. The matrix is the negative of the Jacobian $\partial(P, \overline{\mu}) / \partial(V, N_{m1})$. It is now trivial to see that

$$<(\Delta V)^2>_B = TVK_{T,\bar{\mu}}, \quad <(\Delta N_{m1})^2>_B = T(\partial N_{m1}/\partial \bar{\mu})_{T,P}, \quad (36)$$
$$<\Delta V \Delta N_{m1}>_B = T(\partial V/\partial \bar{\mu})_{T,P},$$

where $K_{T,\bar{\mu}}$ is the isothermal compressibility at fixed $\bar{\mu}$. The volume fluctuation is not given by the usual isothermal compressibility at fixed monomer numbers. Instead, it is given by $K_{T,\bar{\mu}}$. This result is in accordance with Theorem 2. We also note that the monomer number fluctuation is different from the A-ensemble; in particular, it is evaluated at constant $P$ instead of at constant $V$, again in accordance with Theorem 2.

The coupling between the volume and monomer fluctuations gives rise to interesting effects. Using the first of Eq. (8) and Eqs. (32), and (36), we can evaluate the fluctuation in $N_{m2}$ and the cross fluctuation.

$$<(\Delta N_{m2})^2>_B = <(\Delta V)^2>_B + [1 - 2(\partial V/\partial N_{m1})_{T,P}]<(\Delta N_{m1})^2>_B,$$
$$<\Delta N_{m1}\Delta N_{m2}>_B = [(\partial V/\partial N_{m1})_{T,P} - 1]<(\Delta N_{m1})^2>_B, \quad (37a)$$

where we have used the following relation that follows from Eq. (36)
$$<\Delta V \Delta N_{m1}>_B = (\partial V/\partial N_{m1})_{T,P}<(\Delta N_{m1})^2>_B. \quad (37b)$$

Thus, all fluctuations involving monomer numbers can be expressed in terms of at most *two* fluctuations, viz. $<(\Delta N_{m1})^2>_B$ and $<(\Delta V)^2>_B$, which is in accordance with the corollary. For $\Delta V = \Delta N_m$, and $\Delta N_{m1}$, $n' = 1$, which justifies the first two equations in Eq. (36). For the fluctuation in $N_{m2}$, we have $n' = 2$, which justifies the first equation in Eq. (37a). The cross-fluctuations are expressed in terms of self-fluctuations in accordance with Theorem 4. We should stress that all derivatives above are at constant U={$V_0$}. Therefore, $\bar{v}_{01} \equiv (\partial V/\partial N_{m1})_{T,P,V_0}$ in Eqs. (37a,b) is related to the partial monomer volume $\bar{v}_j \equiv (\partial V/\partial N_{mj})_{T,P,N_{mk}}$; the latter requires keeping $N_{mk}$ different from $N_{mj}$ fixed and not $V_0$. However, the two are related:

$$\bar{v}_{01} = (\bar{v}_2 - \bar{v}_1)/(\bar{v}_2 - 1), \quad (38)$$

as is easily checked. The volume and monomer fluctuations in Eq. (37b) are uncorrelated if we have an athermal symmetric blend[11,12] for which $\bar{v}_1 = \bar{v}_2$ must always hold.

We now consider the forward scattering intensity $I(0|b_1,b_2)$, which is given by

$$I(0|b_1,b_2) = b_2^2<(\Delta V)^2>_B + \delta b[\delta b + 2b_2\bar{v}_{01}]<(\Delta N_{m1})^2>_B, \quad (39)$$

where $\delta b = (b_1 - b_2)$. We notice that the intensity ($n' = 2$) breaks into two separate terms, one that depends only on the volume fluctuation and the other that depends on only the monomer number fluctuations in accordance with the corollary. A similar partition also occurs in the C-ensemble, as discussed in the Introduction, and the next section. Such partitions do not imply that the two fluctuations are uncorrelated, as Eq. (37b) clearly indicates. Thus, the partition of the intensity into two parts is *not* a consequence of the statistical independence of density and composition fluctuations. To further clarify this point, let us consider fluctuation correlations in the density $\rho$ and the composition $x = N_{m1}/N_m$ explicitly. It is easy to see that the fluctuations are $\Delta \rho = (1-\rho)\Delta V/V$ and $\Delta x = (\Delta N_{m1} - x\Delta V)/N_m$. Therefore, the cross-correlation in the fluctuations is

$$<\Delta \rho \Delta x>_B = [(1-\rho)/VN_m][\bar{v}_{01}<(\Delta N_{m1})^2>_B - x<(\Delta V)^2>_B], \quad (40)$$

and is not zero. Thus, the two fluctuations are not independent in the B-ensemble. This remains true even for an athermal symmetric blend for which $\bar{v}_{01} = 0$.

We observe that $\Delta N_m = V\Delta\rho/(1-\rho)$. Thus, it does not depend explicitly on the composition fluctuation. We can always write $K_{T,\bar{\mu},V_0} = K_{T,N_{m1},N_{m2}} + K_{\text{diff}}$, where $K_{\text{diff}}$ is the difference between the two compressibilities. This allows us to express $<(\Delta N_m)^2>_B$ as a sum of two terms, first of which contains the regular compressibility $K_{T,N_{m1},N_{m2}}$ and the second one contains $K_{\text{diff}}$. Such a partition does not imply that the first term represents the density fluctuation and the second term the composition fluctuation; $<(\Delta N_m)^2>_B$ has no composition contribution.

### IX. C-ensemble

We have recently investigated this ensemble within the context of the neutron scattering experiments.[28] However, the emphasis was different. The system is closed with respect to the volume; therefore, there is no volume fluctuation. This means that all fluctuation correlations with other fluctuating quantities are identically zero:

$$< \Delta V \Delta Q >_C \equiv 0, \tag{41}$$

where $\Delta Q = \Delta V$, $\Delta T$, $\Delta E$, $\Delta N_{m1}$, $\Delta \mu_{m1}$, $\Delta N_{m2}$ and $\Delta \mu_{m2}$. In this ensemble,

$$R = \Delta\beta\Delta E - \Delta\beta\mu_{m1}\Delta N_{m1} - \Delta\beta\mu_{m2}\Delta N_{m2}, \tag{42}$$

which can be re-expressed in terms of $\Delta T$, $\Delta N_{m1}$ and $\Delta N_{m2}$. We find that

$$R = -(1/T)[(1/T)(\partial E/\partial T)_{V,N_{m1},N_{m2}}(\Delta T)^2 + (\partial \mu_{m1}/\partial N_{m1})_{T,V,N_{m2}}(\Delta N_{m1})^2 \\ + (\partial \mu_{m2}/\partial N_{m2})_{T,V,N_{m1}}(\Delta N_{m2})^2 + 2(\partial \mu_{m2}/\partial N_{m1})_{T,V,N_{m2}}\Delta N_{m1}\Delta N_{m2}]. \tag{43}$$

We note immediately that there is also *no* correlation between the temperature and monomer number fluctuations:

$$<\Delta T \Delta N_{m1}>_C = <\Delta T \Delta N_{m2}>_C = 0. \tag{44}$$

This is in accordance with Eq. (12), which also shows that the chemical potentials are uncorrelated with volume or energy fluctuations. Because of the lack of correlation between temperature and monomer number fluctuations, the temperature fluctuation does not get modified by the presence of monomer number fluctuations and has the same value, see the coefficient of $(\Delta T)^2$ in Eq. (43), as it has in the earlier ensembles considered; see Eqs. (16), (18), (25) and (35). This is in accordance with Theorem 3.

The fluctuations in monomer numbers are correlated due to the cross term in Eq. (43). To calculate monomer number fluctuations, we need to calculate the determinant of the matrix formed by the coefficients in Eq. (43):

$$-(1/T)\begin{bmatrix} (\partial\mu_{m1}/\partial N_{m1}) & (\partial\mu_{m1}/\partial N_{m2}) \\ (\partial\mu_{m2}/\partial N_{m1}) & (\partial\mu_{m2}/\partial N_{m2}) \end{bmatrix}; \tag{45}$$

we have suppressed the variables held fixed in the derivatives for simplicity. The determinant is the Jacobian $(1/T^2)\partial(\mu_{m1},\mu_{m2})/\partial(N_{m1},N_{m2})$. Various fluctuations are now trivial to calculate.

$$<(\Delta N_{m1})^2>_C = T(\partial N_{m1}/\partial \mu_{m1})_{T,V,\mu_{m2}}, \quad <(\Delta N_{m2})^2>_C = T(\partial N_{m2}/\partial \mu_{m2})_{T,V,\mu_{m1}},$$
$$<\Delta N_{m1}\Delta N_{m2}>_C = T(\partial N_{m2}/\partial \mu_{m1})_{T,V,\mu_{m2}}. \quad (46)$$

Both monomer fluctuations are *not* independent due to the cross correlation. We can calculate the fluctuation in $N_m$ (for which $n' = 2$), using $\Delta N_m = \Delta N_{m1} + \Delta N_{m2}$. We square this fluctuation and use Eq. (46).

$$<(\Delta V_0)^2>_C = <(\Delta N_m)^2>_C = T[(\partial N_m/\partial \mu_{m1})_{T,V,\mu_{m2}} + (\partial N_m/\partial \mu_{m2})_{T,V,\mu_{m1}}], \quad (47)$$

where we have used the fact that in this ensemble, $\Delta V_0 = \Delta N_m$. Thus, even though $V$ is fixed, $V_0$ fluctuates. Since the cross-fluctuation can be easily expressed in terms of a self-fluctuation in accordance with Theorem 4, we note that the right-hand side of Eq. (47) contains only two fluctuations in accordance with the corollary. The same is also true of the intensity for which $n' = 2$.

The fluctuations can easily be expressed in terms of derivatives calculated at fixed $P$, rather than at fixed $V$.[23] We need the following identities:

$$<(\Delta N_{m1})^2>_C = \rho_1^2 TVK_{T,N_{m1},N_{m2}} - \rho_1\rho_2\bar{v}_2^2 T/(\partial\mu_{m1}/\partial N_{m2})_{T,P,N_{m1}},$$
$$<(\Delta N_{m2})^2>_C = \rho_2^2 TVK_{T,N_{m1},N_{m2}} - \rho_1\rho_2\bar{v}_1^2 T/(\partial\mu_{m1}/\partial N_{m2})_{T,P,N_{m1}}, \quad (48)$$
$$<\Delta N_{m1}\Delta N_{m2}>_C = \rho_1\rho_2[TVK_{T,N_{m1},N_{m2}} + \bar{v}_1\bar{v}_2 T/(\partial\mu_{m1}/\partial N_{m2})_{T,P,N_{m1}}].$$

We use Eq. (2) for the total monomer density fluctuation. We quote the results:

$$<(\Delta N_m)^2>_C = \rho^2 TVK_{T,N_{m1},N_{m2}} - \rho_1\rho_2 T(\bar{v}_2 - \bar{v}_1)^2/(\partial\mu_{m1}/\partial N_{m2})_{T,P,N_{m1}},$$
$$<\Delta N_m\Delta N_{m1}>_C = \rho\rho_1 TVK_{T,N_{m1},N_{m2}} - \rho_1\rho_2 T(\bar{v}_2 - \bar{v}_1)\bar{v}_2/(\partial\mu_{m1}/\partial N_{m2})_{T,P,N_{m1}}, \quad (49)$$

As shown in Eq. (58) below, the cross-derivative $(\partial\mu_{m1}/\partial N_{m2})_{T,P,N_{m1}}$ is non-positive, which makes the second term in each of the two equations in Eq. (49) positive. Since $V$ does not fluctuate here, $\Delta\rho = (\Delta N_m)/V$ and $\Delta x = (\Delta N_{m1} - x\Delta N_m)/N_m$. We obtain

$$VN_m<\Delta\rho\Delta x>_C = <\Delta N_m\Delta N_{m1}>_C - x<(\Delta N_m)^2>_C, \quad (50a)$$

which reduces to a simple form by the use of Eq. (49):

$$<\Delta\rho\Delta x>_C = -\rho_1\rho_2 T(\bar{v}_2 - \bar{v}_1)\bar{v}/(\partial\mu_{m1}/\partial N_{m2})_{T,P,N_{m1}}, \quad (50b)$$

where we have introduced the average partial monomer volume
$$\bar{v} = x\bar{v}_1 + (1-x)\bar{v}_2.$$

It is now clear that the cross correlation $<\Delta\rho\Delta x>_C$ is not zero. Pearson and Rushbrooke[32] had also obtained a non-vanishing $<\Delta\rho\Delta x>_C$ in a different form. Thus, the density and composition fluctuations are correlated, contrary to the claims in the literature.[22,23,34]

We observe that for an athermal symmetric blend, the second term in each of the equations in Eqs. (48)-(50) vanishes, leaving the first term behind. Thus, for the special case of an athermal symmetric blend, we note that total monomer number fluctuation has only one contribution, which comes from the compressibility. However, this does not imply that there is no composition fluctuation in the system. Moreover, this also does not imply that there is no cross-correlation between the density and the composition fluctuation.

**Fluctuation Theory Breakdown for $f = 0$.** The next issue we wish to handle is the breakdown of the fluctuation theory if $f = 0$. In this case, we find that the coefficient of some of the fluctuations become zero, or the determinant of the matrix vanishes for the case there are fluctuations with cross terms, as happened in the B- and C- ensembles. Consider the C-ensemble, but allow for the remaining fixed extensive quantity $V$ to fluctuate. Now, there will be an additional term $(-\Delta\beta P\Delta V)$ in Eq. (42) and an additional term in $(\Delta P)^2$ in Eq. (43), if we use $\Delta P$ as an expansion variable. There are no cross-terms between $\Delta P$ and number fluctuations. Hence, the number fluctuation contributions give rise to a 2X2 matrix, similar to that in Eq. (45), except that the derivatives in the matrix are evaluated at $T$, $P$ and one of the remaining monomer numbers. However, the determinant of this matrix now vanishes, implying that the fluctuations diverge everywhere in the phase space. While fluctuations in monomer numbers can diverge at critical points, they occur for special values of the fields, not everywhere. The implication of the vanishing determinant is that the fluctuation theory becomes singular and meaningless.

## X. Composition and Density Fluctuations

We now turn to the composition and density fluctuations. We have already considered some trivial aspects of this in previous sections. Here, we wish to make formal analysis to clarify the issue, which in our opinion has been vastly misunderstood. We first express the two monomer numbers in terms of $x$, $\rho$, and $V$.

$$N_{m1} = x\rho V, \; N_{m2} = (1-x)\rho V,$$

from which the fluctuations are found to be

$$\Delta N_{m1} = \rho V\Delta x + xV\Delta\rho + x\rho\Delta V,$$
$$\Delta N_{m2} = -\rho V\Delta x + (1-x)V\Delta\rho + (1-x)\rho\Delta V. \tag{51}$$

Thus,

$$\Delta N_m = V\Delta\rho + \rho\Delta V. \tag{52}$$

We immediately note that the total number fluctuation does *not* depend *explicitly* on the composition fluctuation $\Delta x$. It is determined only by the density and volume fluctuations. Both these fluctuations implicitly depend on the presence of composition fluctuations, as all fluctuations in the model are in general intertwined due to correlations; see Theorem 1 and the corollary. On the other hand, the composition fluctuation always depend *explicitly* on the total number fluctuation since

$$N_m\Delta x = \Delta N_{m1} - x\Delta N_m. \tag{53}$$

Because of this explicit dependence, the cross fluctuation $\Delta\rho\Delta x$ also does not vanish in general:

$$VN_m\Delta\rho\Delta x = (\Delta N_m - \rho\Delta V)(\Delta N_{m1} - x\Delta N_m). \tag{54}$$

For a single component system, $x\equiv 1$ and *no* composition fluctuation is allowed. In this case, there can only be a density fluctuation. This brings about a very important feature of the density fluctuation. The fluctuation $<(\Delta\rho)^2>$ in a *single* component system must remain the same, whether we take a constant volume ensemble or a constant number ensemble. Because of this,

$$<(\Delta N_m)^2>_{T,V,\mu_m}/N_m^2 \equiv <(\Delta V)^2>_{T,P,N_m}/V^2; \tag{55}$$

the quantities in the denominator are the average values. An elegant proof of this is given in Ref. 1. Here, $\mu_m$ is the chemical potential per particle (in this case, monomer). The important point to note is that the two sides of Eq. (55) refer to two *different* ensembles. One should not confuse them with fluctuations in the same ensemble, which can occur simultaneously as in the B-ensemble. We see that there is no $\rho^2$-factor in Eq. (32); moreover, both sides refer to the same ensemble. For a binary mixture, Eq. (55) is not going to be valid, because of the additional composition fluctuation. However, an extension, see Eq. (65), is proven below for a multi-component mixture.

In the A-ensemble, $\Delta V = \Delta N_m = 0$. In the B-ensemble, the volume and total monomer number fluctuations are equal; see Eq. (32). We further recall from Eq. (39) that the intensity breaks into two terms, the first one containing the volume fluctuation and the second term containing the monomer number fluctuation. However, this alone is not a proof that the volume and the number fluctuations or that the density and the composition fluctuation are uncorrelated, as is clear from Eqs. (37b), and (40). Even $\langle \Delta x \Delta N_m \rangle$ are correlated. Thus, the partition of $I(0|b_1,b_2)$ into two distinct terms is not a guarantee that the above cross correlations are absent. However, exactly this argument has been proposed in Ref. 25; see discussion following Eq. (8) there.

In the C-ensemble, the extension of the left-hand side of Eq. (55) is given in Eq. (49). The right-hand side is given in Eq. (19), with $X_{NF}=\{N_{m1}, N_{m2}\}$, which also ensures that $N_m$ is fixed, so there cannot be any composition fluctuation. We note that Eq. (55) fails because of the second term in Eq. (49). In the absence of this term, Eq. (55) would be satisfied.

The reason for this failure is obvious. As is clear from Eq. (52), $\langle (\Delta N_m)^2 \rangle_C$ is solely due to the density fluctuation $\langle (\Delta \rho)^2 \rangle$, since the volume is constant. Therefore, both terms together in Eq. (49) represent the density fluctuation $\langle (\Delta \rho)^2 \rangle$. It is incorrect to claim that just the first term represents $\langle (\Delta \rho)^2 \rangle$. As said above, the first part is $\rho^2 \langle (\Delta V)^2 \rangle_D$ in the D-ensemble, i.e., $T$-$P$-$X_{NF}$ ensemble, see Eq. (19). But the volume fluctuation is absent in the C-ensemble. The remainder still represents a part of $\langle (\Delta N_m)^2 \rangle_C \equiv V^2 \langle (\Delta \rho)^2 \rangle_C$, and is the correction to the density fluctuation, see the Theorem 1, originating from the composition fluctuation, but it surely is not the composition fluctuation itself. This is apparent from Eqs. (53), and (54). Because of the cross correlation in Eq. (54), the density and the composition fluctuations are modified by each other.

It should also be evident from Eq. (51) that to capture composition fluctuation, we should not consider the fluctuation $\langle (\Delta N_m)^2 \rangle$. We might consider the total intensity in Eq. (1) with different scattering lengths $b_1$ and $b_2$. In this case, the $\Delta x$ contributions in Eq. (51) will not cancel, and the total intensity will contain both the density and the composition fluctuations.

It is possible to introduce[2] the two following combinations of number fluctuations
$$\xi = (\bar{v}_1 \Delta N_{m1} + \bar{v}_2 \Delta N_{m2})/V, \quad \xi' = \Delta N_{m1}/N_{m1} - \Delta N_{m2}/N_{m2}, \tag{56}$$
so that $V\xi$ represents a *weighted monomer number fluctuation* with weights $\bar{v}_j$ and $\xi'$ represents relative fractional number fluctuation. We find that

$$<\xi^2>_C = TK_{T,N_{m1},N_{m2}}/V, \qquad <\xi\xi'>_C = 0, \tag{57a}$$

$$<\xi'^2>_C = -T[N_{m1}N_{m2}(\partial\mu_{m1}/\partial N_{m2})_{T,P,N_{m1}}]^{-1}, \tag{57b}$$

as is easily checked from Eq. (48). From Eq. (57b), we conclude that

$$(\partial\mu_{m1}/\partial N_{m2})_{T,P,N_{m1}} \leq 0. \tag{58}$$

It is easy to see that if we introduce the concentration

$$c = N_{m1}/N_{m2},$$

we have $\xi' = \Delta c/c$ and $\xi'$ has a clear physical significance. Indeed, $\Delta x = x(1-x)\xi'$, as is easily seen from Eq. (53). On the other hand, $\xi$ does not have any simple physical interpretation. It has been incorrectly interpreted as representing the density fluctuation at constant composition,[2] as we show below. At constant $c$, we have $\Delta N_1 = c\Delta N_2$, which makes the three fluctuations in Eq. (48) related to each other. We immediately see that for this to be true, we must have

$$(\partial\beta\mu_{m1}/\partial N_{m2})^{-1}_{T,P,N_{m1}} = 0. \tag{59}$$

This can also be deduced from Eq. (57b). From Eq. (48), we find that

$$<(\Delta N_m)^2>_C \equiv V^2 <(\Delta\rho)^2>_C = \rho^2 TVK_{T,N_{m1},N_{m2}}, \quad (c = \text{constant}) \tag{60a}$$

where we have used the identity $V = \bar{v}_1 N_{m1} + \bar{v}_2 N_{m2}$ that yields $\bar{v}_1\rho_1 + \bar{v}_2\rho_2 \equiv 1$. The result is not surprising as there is no composition fluctuation now, which also means that the cross-fluctuations with composition also vanish. However, it is clear from Eqs. (57), and (60a) that the density fluctuation $<(\Delta\rho)^2>_C$ is not the same as $<\xi^2>_C$, though the two are related:

$$<(\Delta\rho)^2>_C = \rho^2 <\xi^2>_C, \quad (c = \text{constant}). \tag{60b}$$

However, Eq. (59) cannot be satisfied in general in the C-ensemble, except possibly at some isolated points. From the Gibbs-Duhem relation, we find that Eq. (59) is equivalent to

$$(\partial\beta\mu_{m1}/\partial N_{m1})^{-1}_{T,P,N_{m2}} = 0; \quad (\partial\beta\mu_{m2}/\partial N_{m2})^{-1}_{T,P,N_{m1}} = 0. \tag{61}$$

These conditions can be satisfied only when $\beta\mu_{m1}$ and $\beta\mu_{m2}$ both diverge to infinity, so that the equilibrium values of $N_{m1}$ and $N_{m2}$ are their extremum values under appropriate conditions. This means that the point where there is no fluctuation in $\xi'$ is a Nernst point. It is also evident from Eq. (9) that in this limit, the free volume must vanish, as $N_m$ must be its maximum possible value, which is $V$. Thus, in this limit, the compressibility must also vanish. This is consistent with the fact that $\beta P \to \infty$ in this limit, as the adimensional Gibbs free energy per monomer $\beta G \equiv x\beta\mu_{m1} + (1-x)\beta\mu_{m2} \equiv \beta E - S + \beta P/\rho \to \infty$. Thus, we observe from Eq. (45) that

$$<(\Delta N_{m1})^2>_C = <(\Delta N_{m2})^2>_C = 0, \quad (c=\text{constant}), \tag{62}$$

and we conclude that the vanishing of any one number fluctuation leads to the vanishing of all number fluctuations. This is a remarkable result. As a matter of fact, it is easy to convince oneself that the above conditions for non-fluctuating $\xi'$ are equivalent to requiring $T = 0$. Therefore, a constant $c$ can only occur at $T = 0$ and is not to be taken seriously. Moreover, Eq. (62) clearly shows that the fluctuation in $\xi$ is indirectly affected by the fluctuation in $\xi'$, even though the two are statistically independent. This is a consequence of thermodynamics. The fact is that the fluctuations are uniquely determined by $T$, $\mu_{m1}$, $\mu_{m2}$ along with $V$, and cannot be changed without changing the thermodynamic variables. Thus, demanding that the fluctuation in Eq. (57b) vanish requires $T = 0$, which in turn modifies the fluctuation in $\xi$.

There is another aspect of the fluctuation $\xi$ that requires explanation. The fluctuation in the C-ensemble requires considering a region $\mathcal{R}$ that has a fixed volume $V$, but the numbers of monomers in this region fluctuate. As they fluctuate, the total volume is not allowed to change. Thus, the partial monomer volume $\bar{v}_1$ and $\bar{v}_2$ must fluctuate in such a way that

$$\bar{v}_1 \Delta N_{m1} + \bar{v}_2 \Delta N_{m2} = -(\Delta \bar{v}_1 N_{m1} + \Delta \bar{v}_2 N_{m2}). \tag{63}$$

As said above, the $\xi$ on the left-hand side represents a weighted monomer number fluctuation in the C-ensemble. Can we think of it as representing a *hypothetical* change in the volume due to number fluctuations? The hypothetical fluctuation cannot occur at constant $T$ and $P$, if both particle numbers undergo fluctuations, since this will require $f = 0$ and will violate our fundamental constrain $f \geq 1$. If only one of the numbers fluctuates and the other one remains fixed, then the concentration cannot remain fixed. Thus, the hypothetical fluctuation is not realistic. The right-hand side of Eq. (63) also cannot be given a physical interpretation in the C-ensemble. However, it can be given a physical significance in the $T$-$P$-$N_{m1}$-$N_{m2}$ ensemble, i.e., the D-ensemble. In this ensemble, volume fluctuation will give rise to partial monomer volume fluctuations and will be given by the negative of the right-hand side of Eq. (63). The average volume fluctuation in the D-ensemble is given in Eq. (19), which is identical to the fluctuation in left-hand side of Eq. (63) in the C-ensemble as is clear from Eq. (57). As said earlier, $\xi$ represents a weighted number fluctuation. Let us introduce the following weighted sums:

$$\begin{aligned}\tilde{N}_m &\equiv \rho \bar{v}_1 N_{m1} + \rho \bar{v}_2 N_{m2} \equiv N_m, \\ \tilde{\Delta} N_m &\equiv \rho \bar{v}_1 \Delta N_{m1} + \rho \bar{v}_2 \Delta N_{m2} \equiv \rho V \xi,\end{aligned} \tag{64}$$

with respect to the weights $\rho \bar{v}_j$. The weighted average fluctuation $\tilde{\Delta} N_m$ should not be confused with the fluctuation $\Delta \tilde{N}_m$ in the weighted average $\tilde{N}_m$. We observe that

$$\xi = \tilde{\Delta} N_m / \tilde{N}_m.$$

Thus, we conclude that

$$<(\tilde{\Delta} N_m)^2>_C / N_m^2 \equiv <(\Delta V)^2>_D / V^2 \equiv T K_{T, N_{m1}, N_{m2}} / V. \tag{65}$$

This result for a two-component compressible system is similar to the result in Eq. (55) for a compressible one-component system and is valid regardless of whether $c$ is constant or not. The quantity $K_{T, N_{m1}, N_{m2}}$ represents the volume fluctuation in the D-ensemble, but does not represent the density fluctuation in the C-ensemble. In addition, as said above, $\xi$ has no simple physical significance in the C-ensemble.

The extension of Eq. (62) to multi-component mixture containing $r$ species is obvious, and we will only quote the results.[2] We introduce

$$\xi \equiv \sum_{j=1}^{r} \bar{v}_j \Delta N_{mj}/V, \quad \xi^{(j)} \equiv \Delta N_{mj}/N_{mj} - \Delta N_{mr}/N_{mr},$$

$$\tilde{N}_m \equiv \sum_{j=1}^{r} \rho \bar{v}_j N_{mj} \equiv N_m, \quad \tilde{\Delta} N_m \equiv \sum_{j=1}^{r} \rho \bar{v}_j \Delta N_{mj} \equiv N_m \xi. \quad (66)$$

As shown in Ref. 2, Eq. (57a) remains valid even for multi-component mixtures. Thus, there is no cross-fluctuation between $\xi$ and $\xi^{(j)}$. However, there are cross fluctuations among $\xi^{(j)}$'s. Again, using an appropriate linear transformation from $\xi^{(j)}$ to $\xi'^{(j)}$, the cross-fluctuations can be removed among $\xi'^{(j)}$'s in accordance with Theorem 1. Using the multi-component extension of Eq. (63), we have $\xi = -\sum_{j=1}^{r} \Delta v_j N_{mj}$. In the D-ensemble, the right-hand side represents the negative of the volume fluctuation. Thus, the volume fluctuation in the D-ensemble is given by Eq. (19). Thus, Eq. (65) remains valid even for multi-component system.

## XI. Discussion and Conclusions

In this review, we have discussed the general framework for describing statistical fluctuations in a given fixed region $\mathcal{R}$ of a thermodynamic system. The thermodynamic limit is obtained as the size of the region diverges. Thus, at least one extensive quantity is required to specify the size of the system. Consequently, the thermodynamic degree $n$ of the ensemble must be strictly less than the maximum allowed thermodynamic extensive quantities $d$ in the system. This requirement must be obeyed in all ensembles. Corresponding to each extensive quantity $X$, there exists a field $Y$ or its related analog $y$. We have carefully investigated, and calculated, fluctuations in various $X$'s and $Y$'s or $y$'s in various ensembles using the statistical mechanical approach outlined by Landau.[1] As expected, the fluctuations vary from ensemble to ensemble.

We have explicitly considered the case in which all monomers had the same hardcore volume $v_0$. The extension to different hardcore volume $v_j$ is not hard. We merely introduce new quantities $N'_{mj} = N_m v_j$, and $\mu'_{mj} = \mu_{mj}/v_j$, so that the product $N'_{mj}\mu'_{mj} = N_{mj}\mu_{mj}$ remains unchanged. All we must do now is to replace $N_{mj}$ and $\mu_{mj}$ in various formulas by $N'_{mj}$ and $\mu'_{mj}$. No other changes have to be made. However, since the intensity in Eq. (1) is from the fluctuations in unprimed quantities, we must replace $b_j$ by $b'_j = b_j/v_j$, and use primed quantities in Eq. (1).

We have proved four general theorems and a corollary for general statistical fluctuations. These theorems prove extremely useful. We have shown that the number of statistically independent fluctuations is equal to $n$, the thermodynamic degree of the ensemble. The fluctuation in any quantity can have at most $n$ fluctuating contributions. This means, there can be at most two different contribution to the total intensity in the SANS experiment on a binary blend. Cross-correlations among extensive quantities can always be expressed in terms of self-correlations. The field variables only couple to their conjugate extensive quantities but not to other extensive quantities. Thus, $T$ and/or $P$ do not couple to number fluctuations. The number fluctuations are correlated. We have established that the volume and monomer number fluctuations are not really decoupled;

see Eq. (37b). Moreover, the density and composition fluctuations are not decoupled as seen from Eq. (40). This is in the B-ensemble. In the C-ensemble, we again see from Eq. (50) that the density and composition fluctuations are coupled. Recently, Benoit and coworkers[38] have finally come to agree with this conclusion. Moreover, the density fluctuation, which is obtained by dividing $<(\Delta N_m)^2>_C$ by $V^2$, is given by both terms in Eq. (50). Thus, the first term alone *cannot* represent the density fluctuation as has been suggested in the literature. In the special case of an athermal symmetric blend, the total number fluctuation $<(\Delta N_m)^2>_C$ contains only the first term; the second contribution is absent. This does not mean that there is no composition fluctuation in the system. It is still present for the athermal symmetric blend. It just does not contribute to $<(\Delta N_m)^2>_C$. Indeed, the volume and density fluctuations will *not* decouple in a multi-component system.

We show that while the theory of fluctuations is derived under the assumption of small fluctuations, the results have a much wider applicability. For example, one can apply them near critical points where fluctuations diverge. Several applications of the theorems and corollary are discussed in the review.

The fluctuation theory is equally applicable to equilibrium states and to stationary metastable states. It is usually believed that one can produce, at least hypothetically, stationary metastable states in the form of ideal glass by carefully preparing the sample as slowly as possible, ensuring all the time that the equilibrium crystal phase does not nucleate. The ideal glass is an inactive phase with zero heat capacity. We consider the consequences of vanishing susceptibilities, like the heat capacity, which define Nernst points. We prove a generalized Nernst Theorem.[5] As a consequence of this theorem, we argue that when the heat capacity vanishes, it causes *anomalous fluctuations* in the temperature and that temperature loses any physical significance. Thus, the ideal glass has no unique temperature. Consequently, it can be brought into thermal equilibrium with any physical system at any temperature. This is evidently absurd. Thus, we are forced to conclude that stationary metastable states of vanishing heat capacity like the ideal glass is impossible in Nature. In other words, no glassy phase can have a stationary limit. Its time-dependence can never be made to disappear. It should be mentioned that there exist many statistical mechanical models, which have inactive low-temperature phase. However, all these models are *not* realistic. Thus, while the existence of such phases violates the zeroth law of thermodynamics, it poses no problem for applying thermodynamics to real systems. However, if ideal glass exists in Nature, this certainly causes the zeroth law to fail for real systems. Hence, we are forced to conclude that ideal glass cannot exist in Nature.

We have shown that the vanishing of any fluctuation leads to the vanishing of all fluctuations simultaneously, which is a quite remarkable result. The fluctuation $\xi$ or $\tilde{\Delta}N_m$ has no physical significance in the C-ensemble. We have also shown that considering fluctuation $\xi$ at constant $c$ makes *no* sense, except at absolute zero. The issue does not arise in a single component system, where Eq. (55) is always obeyed. An extension of this result for a multi-component mixture is obtained in Eq. (65). Moreover, we have also shown that even if two fluctuations are statistically uncorrelated, they still influence each other because of thermodynamics, as the discussion on the behavior of $\xi$ has clearly demonstrated when the composition was fixed, i.e. when $\xi'$ was zero.

# References

1. L. D. Landau, E. M. Lifshitz and L. P. Pitaevskii, *Statistical Physics*, Part 1 (Pergamon, Oxford, 1980).
2. T. L. Hill, *Statistical Mechanics*: *Principles and selected applications*, McGraw-Hill, New York (1957).
3. H. B. Callen, *Thermodynamics and Introduction to Thermostatistics,* 2$^{nd}$ Ed. (Wiley, New York, 1985).
4. P. T. Landsberg, *Thermodynamics and Statistical Mechanics* (Oxford Univ. Press, Oxford, 1978).
5. P. D. Gujrati, *Phys. Lett*. **151A**, 375 (1990).
6. P. D. Gujrati, S. S. Rane and Andrea Corsi, to be published.
7. P. D. Gujrati, to be published.
8. C. Domb, M. Green and J. Lebowitz, eds. *Phase Transitions and Critical Phenomena* (Academic Press, New York).
9. S. Wolf and V. Kresin, eds., *Novel Superconductivity* (Plenum, New York, 1987).
10. R. M. Wald, *General Relativity* (Univ. of Chicago Press, Chicago, 1984).
11. K. S. Thorne, in *Highlights of Modern Astrophysics*, eds. S. L. Shapiro and S. A. Teukolosky (Wiley, New York, 1986).
12. R. Balian, R. Maynard and G. Toulouse, eds. *Ill-Condensed Matter* (North-Holland, Amsterdam, 1979).
13. M. A. Moore and C. A. Wilson, *J. Phys.* **A**13, 3501 (1980).
14. R. B. Griffiths and P. D. Gujrati, *J. Stat. Phys.* 30, 563 (1983).
15. M. Goldstein and R. Simha, eds. *Ann. N. Y. Acad. Sci.* 279 (1976).
16. J. Jäckle, Models of the glass transition, *Rep. Prog. Phys.* 49 171 (1986).
17. W. Nernst, *Thermodynamics and Chemistry* (New York, 1907); *Die Theoretischen und Practishen Grundlagen des neuen Wärmesätzes* (Halle, 1918).
18. J. G. Kirkwood and R. J. Goldberg, *J. Chem. Phys.*, **18**, 54 (1950).
19. W. H. Stockmayer, *J. Chem. Phys.*, **18**, 58 (1950).
20. J. des Cloizeaux and G. Jannink, *Physica* **102A**, 1206 (1980).
21. H. Benoit, M. Benmouna and W. Wu, *Macromolecules* **23**, 1511 (1990).
22. H. Benoit, *Polymer* **32**, 579 (1990).
23. J. S. Higgins and H. C. Benoit, *Polymers and Neutron Scattering* (Clarendon, Oxford, 1994).
24. J.-F. Joanny, H. Benoit and W. H. Stockmayer, *Macromol. Symp.* **121**, 95 (1997).
25. J.-F. Joanny and H. Benoit, *Macromolecules* **30**, 3704 (1997).
26. P. D. Gujrati, *J. Chem. Phys.* **112**, 4806 (2000).
27. P. D. Gujrati, *Bull. Am. Phys. Soc*. V15.005 (2001).
28. Sagar Rane and P. D. Gujrati, submitted for publication.
29. H. C. Brinkman and J. J. Hermans, *J. Chem. Phys.* **17**, 574 (1949).
30. W. H. Stockmayer, *J. Chem. Phys.* **18**, 58 (1950).
31. J. G. Kirkwood and F. P. Buff, *J. Chem. Phys.* **19**, 774 (1951).
32. F. J. Pearson and G. S. Rushbrooke, *Proc. Roy. Soc. Edin*. **A64**, 305 (1957).
33. J. Des Cloizeaux and G. Jannink, *Physica* **102A**, 120 (1980).

34. J. K. Taylor, P. G. Debenedetti, W. W. Grasley and S. K. Kumar, *Macromolecules*, **30**, 6946(1997).
35. K-K Han and H. S. Son, *J. Chem. Phys.* **115**, 7793 (2001).
36. A. Einstein, *Investigations on the theory of the Brownian movement*, Dover (1936).
37. E. M. Purcell and R. V. Pound, *Phys. Rev.* **81**, 279 (1951); N. F. Ramsey, *Phys. Rev.* **103**, 20 (1956).
38. A. Z. Akcasu, G. Jannink, and H. Benoit, *Eur. Phys. J.* E**8**, 315 (2002).